\documentclass[a4paper,prd,twocolumn,superscriptaddress]{revtex4}

\begin{document}

\newcommand{\be}{\begin{equation}}
\newcommand{\ee}{\end{equation}}
\newcommand{\bq}{\begin{eqnarray}}
\newcommand{\eq}{\end{eqnarray}}
\newcommand{\bsq}{\begin{subequations}}
\newcommand{\esq}{\end{subequations}}
\newcommand{\bc}{\begin{center}}
\newcommand{\ec}{\end{center}}

\newcommand {\R}{{\mathcal R}}
\newcommand{\al}{\alpha}

\title{Cosmic Numbers: A Physical Classification for Cosmological Models}

\author{P.P. Avelino}
\email[Electronic address: ]{pedro@astro.up.pt}
\affiliation{Centro de Astrof\'{\i}sica da Universidade do Porto, R. das
Estrelas s/n, 4150-762 Porto, Portugal}
\affiliation{Departamento de F\'\i sica da
Faculdade de Ci\^encias da Universidade do
Porto, Rua do Campo Alegre 687, 4169-007, Porto, Portugal}
\author{C.J.A.P. Martins}
\email[Electronic address: ]{C.J.A.P.Martins@damtp.cam.ac.uk}
\affiliation{Centro de Astrof\'{\i}sica da Universidade do Porto, R. das
Estrelas s/n, 4150-762 Porto, Portugal}
\affiliation{Department of Applied Mathematics and Theoretical Physics,
Centre for Mathematical Sciences,\\ University of Cambridge,
Wilberforce Road, Cambridge CB3 0WA, United Kingdom}
\affiliation{Institut d'Astrophysique de Paris, 98 bis Boulevard Arago,
75014 Paris, France}

\date{15 October 2002}

\begin{abstract}
We introduce the notion of the cosmic numbers of a cosmological
model, and discuss how they can be used to naturally classify models
according to their ability to solve some of the problems of the
standard cosmological model.
\end{abstract}
\pacs{98.80-k,98.80.Es,98.80.Hw}
\keywords{}
\preprint{DAMTP-2002-119}
\maketitle

The idea of possible time and space variations of the
fundamental `constants' of nature has attracted a lot of interest
in recent times.
This has been powered, on the observational side, by the
observational hints for possible variations of the fine structure
constant,  $\alpha$ \cite{Webb}, and the ratio of the proton and electron
masses, $\mu=m_p/m_e$ \cite{Ivanchik}, detected by spectroscopy of low-redshift
astrophysical sources. A further, arguably less
robust hint also comes from the Oklo natural reactor \cite{Fujii},
while the Cosmic Microwave Background and Big Bang Nucleosynthesis
provide relatively weaker bounds, though at much higher
redshifts \cite{Peak,Avelino,Martins}. Further constraints are also
discussed in \cite{Uzan}.

On the theoretical side there have been several claims
that some of the major problems of the standard cosmological
model \cite{KTurner} could be solved in a
varying $\alpha$ theory if $\alpha$ was
smaller in the past, or alternatively in so-called varying speed of light
(VSL) theories \cite{Moffat,Albrecht,BM1,BM2}
or in bi-metric theories \cite{Bassett,Adiabatic,Gaussian}.
At a different level, varying fundamental constants are ubiquitous in theories
with additional spacetime dimensions \cite{Essay}.

There has also been some controversy on whether or not
it is always possible to identify which of the fundamental `constants'
are varying \cite{Duff}. It is worth pointing out
that some of this this discussion
is rather `academic' in scope, often relying on rather contrived
interpretations of particular definitions or concepts.
Here we aim to clarify this debate, in the spirit of \cite{Previous}
by providing a physical
classification of cosmological models according to the behaviour
of their characteristic cosmic numbers (which will be
defined below).
From a practical (experimental) point of view, it should be obvious
that experiments can only measure
dimensionless combinations of fundamental parameters, despite some
attempts to classify theories as varying electric charge ($e$) or
varying speed of light ($c$)
depending on simplicity arguments \cite{CorE,Davies}.
This means that when measuring
a dimensional parameter \textit{one should always specify the choice of
units in which it is measured}. 

Let us now focus on cosmology, and assume that the cosmological principle
holds. Then there is always a class of (co-moving) observers for which
the universe is homogeneous and isotropic. This space-time is therefore
described by a Friedmann-Robertson-Walker metric,
$$
ds^2=-c^2dt^2+a^2(t)\left(\frac{dr^2}{1-kr^2}+r^2d\Theta^2_2\right)\,,
$$
regardless of any possible modifications to the dynamics of our
theory of gravity. We expect that such modifications will
appear naturally in the context of the possible variations of the
fundamental `constants' of nature.

This means that, quite generally, cosmology
provides us with two `natural' units: one unit of length,
\textit{the curvature scale}, $\ell_c\equiv a |k|^{-1/2}$,
and one unit of time, \textit{the Hubble time},
$H^{-1} \equiv a/(da/dt)$. Hence, the natural way (in a cosmological
sense) of measuring the speed of light in a given cosmological
model is to express it in these units. We shall call this the
\textit{expansion number}, and define it as
$$
C_e\equiv \frac{c |k|^{1/2}}{a H}\,.
$$
Note that this is a dimensionless quantity.
With this definition, we immediately see that `varying cosmic speed'
(VCS) theories are generic \cite{Previous}.
For example, in the standard cosmological model one has
$C_e \propto t^{1/2}$ during the radiation dominated
era, $C_e \propto t^{1/3}$ during the matter dominated era,
$C_e =const.$
during a curvature dominated era, and
$C_e \propto \exp \left(-Ht\right)$ if the
cosmological constant dominates.

One can easily show with all generality (that is,
assuming only the cosmological principle)
that the resolution of the horizon and
flatness problems depends naturally on the behaviour of the expansion
number $C_e$, rather than on the behaviour of $c$
in units of $e^2/\hbar$ (in which case it would be related to
the variation of $\alpha$). Before discussing the general case,
we notice as an aside that the flatness problem is in fact
trivially solved if the \textit{standard} Friedmann equation holds, since in
that case the expansion number defined above reduces to
$$
C_{e,st}=\left|\Omega-1\right|{}^{1/2}\,.
$$

However, the main point we wish to clarify in this letter is that 
regardless of the explicit expression for $C_e$ (which will of course
be model-dependent), it will still be true that the behaviour of
this parameter will determine whether or not the horizon and
flatness problems can be solved.

Indeed, a solution to the flatness problem depends 
only on the behaviour of $C_e$, independently of any modifications 
to the Friedmann equation. This is easy to understand given that the flatness 
problem can be solved if our theory explains why the current 
Hubble radius, $c_0 H^{-1}_0$, is much smaller than (or of the same order 
of) the curvature scale, $a_0|k|^{-1/2}$. On the other hand, a solution to the 
horizon problem can be achieved by having a period in the history of the 
Universe in which the scale factor, $a$, grows faster than the Hubble radius, 
$c H^{-1}$. In both cases, we assume that the Cosmological 
Principle holds for a region with size at least equal to the initial Hubble 
radius. Note that the same is required in inflationary models: assuming that
General Relativity holds, the vacuum energy density which drives
inflation must not be very inhomogeneous over a region of several
Hubble radii if inflation is to begin.

In summary, what this means is that
both the horizon and flatness problems can be solved if
$$
\frac{d}{dt} C_e < 0\,,
$$
for a suitable (that is, large enough)
period of time in the early Universe.
We call these decreasing cosmic speed (DCS) theories,
as opposed to increasing cosmic speed (ICS) ones. Therefore
we see that both the standard inflationary paradigm, the
deflationary phase in the pre-Big-Bang scenario  \cite{Gasperini}
and some of the so-called
VSL theories are examples of DCS theories, and they are therefore
capable of solving these cosmological problems.
On the other hand, the standard cosmological
model is an example of an ICS theory, and hence can not solve them.

We emphasize that this classification is entirely
independent of what is happening to
the dimensionful quantities in the model---an obvious point, since one
can always change this behaviour by a suitable re-definition of our
units of measurement (for length, time, and energy). But more to the
point, this is
\textit{also generically independent} of the behaviour of 
dimensionless particle physics quantities in the model.

It is also worth stressing that the existence of a period
where the above condition holds is \textit{necessary} but
may not be sufficient for these problems to be
solved. While satisfying the above equation, the universe will
be getting closer to flatness, but it could for example
re-collapse before such a period begins. This is of course
a well known problem in other contexts, such as inflation
in closed universes.

Other cosmic numbers can be similarly defined. For example, the
\textit{rotation number},
$$
C_r\equiv w/H\,,
$$
and the \textit{shear number},
$$
C_s\equiv \sigma/H\,,
$$
are dimensionless measures
of the angular velocity of the universe, $w$, and the cosmic
shear, $\sigma$, respectively. The
corresponding rotation and shear problems are related to
the un-natural smallness of these numbers \cite{Bunn}.
Inflation is of course an example of
a theory where these problems can be solved.

Hence the message so far is that the cosmological problems can not in
general be solved by an arbitrary
modification of dimensionless particle physics parameters. In general a 
solution to these problems will require an appropriate modification to 
dimensionless variables involving various cosmological quantities.

On the other hand, there are `hybrid' problems whose solution will 
involve combinations of cosmological quantities
and the so-called `fundamental constants of nature'.
For example, explaining why the Hubble time is much larger than the
Planck time requires a period where
$$
\frac{d}{dt}\left(\frac{G{\hbar}H^2}{c^5}\right)< 0\,.
$$
Notice that this involves not only Planck's constant but also Newton's
constant ($G$), and hence the scale of gravity, which can in some
cases have a non-trivial dynamics (brane world models being a
topical example). Furthermore, this also means that the flatness
and oldness problems are in general \textit{different} problems,
and may therefore be solved in some theories by quite different
mechanisms. Only if the \textit{standard} Friedmann equation
is valid do they become the same.

Finally, we note that there are other problems of the standard cosmological
model (such as the seed fluctuations and cosmological constant problems)
that can not be satisfactorily solved in this way. Rather, their solution
is strongly dependent on the detailed 
dynamics of the specific model in question, and therefore no generic
classification can be given.

\bibliography{light}

\end{document}